\documentclass[10pt,twocolumn]{article}
\usepackage[top=1.9cm, bottom=1.9cm, left=1.9cm, right=1.9cm]{geometry}
\usepackage{times}  %
\usepackage[sort&compress,numbers]{natbib}  %
\usepackage[hyphens]{url}
\usepackage{graphicx}  %
\usepackage[keeplastbox]{flushend}
\usepackage[hyphens]{url}  %
\usepackage{graphicx} %
\usepackage{tikzsymbols}
\newcommand{\revision}[1]{\textcolor{black}{#1}}
\newcommand{\newrevision}[1]{\textcolor{black}{#1}}
\newcommand{\descr}[1]{\smallskip\noindent\textbf{#1}}

\usepackage{sectsty}
\sectionfont{\bfseries\Large\raggedright}

\usepackage{url}                                %
\usepackage{array,multirow,graphicx,adjustbox}  %
\usepackage{booktabs}                           %
\usepackage[utf8]{inputenc}
\usepackage[ruled,algosection,noend,linesnumbered]{algorithm2e}
\usepackage{float}
\usepackage{paralist}
\usepackage{amsmath}
\usepackage{amsfonts}
\usepackage{xcolor}
\usepackage{csquotes} 
\usepackage{tabularx}
\usepackage{makecell}
\usepackage{pbox}
\usepackage[hang,flushmargin]{footmisc}
\usepackage{flushend}
\usepackage{xspace}
\usepackage{multicol, lipsum}
\usepackage[T1]{fontenc}

\usepackage{xcolor}
\usepackage{booktabs}
\usepackage{graphicx}
\usepackage{paralist}
\usepackage[small,labelfont=bf]{caption}
\usepackage{subcaption}
\let\oldbibliography\thebibliography
\renewcommand{\thebibliography}[1]{%
  \oldbibliography{#1}%
  \setlength{\itemsep}{2pt}%
}

\usepackage[hang,flushmargin]{footmisc}

\usepackage[compact]{titlesec}
\titlespacing*{\section}{0pt}{*2.5}{2.5pt}
\titlespacing*{\subsection}{0pt}{*2}{2pt}
\usepackage{xspace}

\makeatletter
\def\url@leostyle{%
  \@ifundefined{selectfont}{\def\UrlFont{}}%
  {\def\UrlFont{}}%
}
\makeatother
\usepackage[hyphenbreaks]{breakurl}

\usepackage[bookmarks=true, bookmarksnumbered=true, colorlinks=true, linkcolor=linkcol, citecolor=citecol, urlcolor=urlcol, hypertexnames=true]{hyperref}

\definecolor{darkgreen}{RGB}{0, 100, 0}
\definecolor{linkcol}{rgb}{0.3,0,0}
\definecolor{citecol}{rgb}{0.3,0,0}
\definecolor{urlcol}{rgb}{0.3,0,0}

\makeatletter
\def\url@leostyle{%
  \@ifundefined{selectfont}{\def\UrlFont{\small}}%
  {\def\UrlFont{}}%
}
\makeatother
\begin{document}

\title{\bf Dissecting the Meme Magic: Understanding Indicators\\ of Virality in Image Memes\footnote{To appear at the 24th ACM Conference on Computer-Supported Cooperative Work and Social Computing (CSCW 2021).}}

\author{Chen Ling$^\dag$, Ihab AbuHilal$^\ddag$, Jeremy Blackburn$^\ddag$, Emiliano De Cristofaro$^\mp$,\\ Savvas Zannettou$^\diamond$, and Gianluca Stringhini$^\dag$\\[0.5ex]
\normalsize $^\dag$Boston University, $^\ddag$Binghamton University, $^\mp$University College London, $^\diamond$Max Planck Institute for Informatics\\
\normalsize ccling@bu.edu,  iabuhil1@binghamton.edu, jblackbu@binghamton.edu,\\[-0.5ex]
\normalsize e.decristofaro@ucl.ac.uk, szannett@mpi-inf.mpg.de, gian@bu.edu
}

\date{}
\maketitle

\begin{abstract}

Despite the increasingly important role played by image memes, we do not yet have a solid understanding of the elements that might make a meme go viral on social media.
\newrevision{In this paper, we investigate what visual elements distinguish image memes that are highly viral on social media from those that do not get re-shared, across three dimensions: composition, subjects, and target audience.}
Drawing from research in art theory, psychology, marketing, and neuroscience, we develop a codebook to characterize image memes, and use it to annotate a set of 100 image memes collected from 4chan's Politically Incorrect Board (/pol/).
\newrevision{On the one hand, we find that highly viral memes are more likely to use a close-up scale, contain characters, and include positive or negative emotions.
On the other hand, image memes that do not present a clear subject the viewer can focus attention on, or that include long text are not likely to be re-shared by users.} 

We train machine learning models to \newrevision{distinguish between image memes that are likely to go viral and those that are unlikely to be re-shared}, obtaining an AUC of 0.866 on our dataset.
We also show that \newrevision{the indicators of virality identified by our model can help characterize the most viral memes posted on mainstream online social networks too}, as our classifiers are able to predict 19 out of the 20 most popular image memes posted on Twitter and Reddit between 2016 and 2018.
Overall, our analysis sheds light on what indicators characterize \newrevision{viral and non-viral visual content online}, and set the basis for developing better techniques to create or moderate content that is more likely to catch the viewer's attention.
\end{abstract}

\section{Introduction}
\label{sec:introduction}

Images play an increasingly important role in the in the way people behave and communicate on the Web, including 
emojis~\cite{barbieri2016cosmopolitan,miller2018see,zhou2017goodbye}, GIFs~\cite{bakhshi2016fast,miltner2017never}, and memes~\cite{highfield2016instagrammatics,xie2011visual,zannettou2018origins}.
Image memes have become an integral part of Internet culture, as users create, share, imitate, and transform them. %
They are also %
used to further political messages and ideologies. %
For instance, the Black Lives Matter movement~\cite{jackson2016ferguson} has made extensive use of memes, often as response to racism online or to gather broader support~\cite{leach2017social}.
At the same time, disinformation actors routinely exploit memes to promote false narratives on political scandals~\cite{mcnair2017fake} and weaponize them to spread propaganda as well as manipulate public opinion~\cite{dupuis2019spread,zannettou2019characterizing,zannettou2018origins, williamsdon}.

Compared to textual memes, image memes are often more ``succinct'' %
and, owing to a higher information density, possibly more effective.
Despite their popularity and impact on society, however, we do not yet have a solid understanding of what factors contribute to an image meme going viral. 
Previous work mostly focused on measuring how textual content spreads~\cite{berger2012makes,cataldi2010emerging,suh2010want}, e.g., in the context of commenting~\cite{jamali2009digging}, hashtags~\cite{yang2012we}, online rumors~\cite{shao2016hoaxy}, and news discussion~\cite{lerman2010using,leskovec2009meme}.
These approaches generally look at groups of words %
that are shared online and trace how these spread.

In this paper, we set out to understand the effect that the {\em visual elements} of an image meme play in that image going viral on social media.
We argue that a popular image meme itself can be considered a successful visual artwork, created, spread, and owned by users in a collaborative effort. 
Thus, we aim to understand if image memes that \newrevision{become highly viral} present the same components of successful visual artworks, which can potentially explain why they attract the attention of their viewers. 
\newrevision{At the same time, we want to understand whether images that are poorly composed fail in catching the attention of viewers and are therefore unlikely to get re-shared.}
Moreover, we investigate if the subjects depicted in an image meme, as well as its target audience, can have an effect on its virality.
Overall, drawing from research in vision, aesthetics, neuroscience, communication, marketing, and psychology, we formulate three research hypotheses vis-\`a-vis influence in virality.

\descr{\em RH1: Composition.}
The theory of visual arts has constructed a comprehensive framework to characterize works of art and how  viewers process them and react to them. %
For example, a high-contrast color portrait of a well-depicted character catches viewers' attention at first sight~\cite{pieters2004attention}.
We argue that image memes may attract people's attention in the same way as works of art, and to investigate such hypothesis we use definitions and principles from the arts~\cite{collingwood1958principles}.
Although aesthetics may contribute in a minor way to the success of an image meme, these composition features might set the basis for an image meme to go viral.
Put simply, our hypothesis is that an image meme that is poorly composed is unlikely \newrevision{to be re-shared and become viral}.

\descr{\em RH2: Subject.} 
Our second hypothesis is that the subject depicted in an image meme has an effect on the likelihood of the meme going viral.
Previous research showed that the attention of viewers is attracted by the faces of characters~\cite{buswell1935people}, and that the emotions portrayed in images capture people's attention~\cite{vuilleumier2002facial}.
\newrevision{Therefore, we hypothesize that images that do not have a clearly defined subject do not catch the attention of viewers and are not re-shared.}
\descr{\em RH3: Target Audience.}
Finally, we argue that the target audience has an effect on virality;
specifically, our intuition is that the audience's understanding of a meme inherently impacts their choice to re-share it. %
Thus, memes that require specific knowledge to be fully understood might only engage a smaller number of people and \newrevision{will unlikely become highly viral}.

\descr{Methodology.}
To investigate our three research hypotheses, we follow a mixed-methods approach.
We start by selecting a set of viral and non-viral image memes that were posted on 4chan's Politically Incorrect Board (/pol/), using the computational pipeline developed by Zannettou et al.~\cite{zannettou2018origins}.
We choose /pol/ as previous work showed it is particularly successful in creating memes that later become part of the broader Web's ``culture''~\cite{hine2017kek,nagle2017kill}.
These images serve as a basis to identify visual cues that potentially describe viral and non-viral memes, and allow us to study our research hypotheses. 
Analyzing this set of images and reviewing research from a number of fields, we develop a codebook, \revision{detailed in Section~\ref{sec:framework}},  to characterize the visual cues of an image meme, and identify nine elements that potentially contribute to the virality of a meme. 
In the rest of the paper, we will refer to these elements as {\em features}.

We then have six human annotators label 100 images (50 viral and 50 non-viral) according to the codebook.
Our features can be easily understood by humans, as we measure high inter-annotator agreement, and are discriminative of the viral/non-viral classes.
Finally, we train a classifier to distinguish between viral and non-viral image memes on the labeled dataset.

\descr{Main Results.} Overall, our study yields the following main findings:
\begin{enumerate}
\item %
Our data shows that composition (RH1) does play an important factor in determining which image memes go viral; for instance, images where the subject is taking most of the frame are more likely to belong to the viral class.
As for RH2, we show that image memes that contain characters as subject are indeed more likely to go viral; in particular, this is true for characters that express a positive emotion (for example through their facial expression).
For RH3, we do not find a significantly higher chance for image memes that do not require background knowledge to go viral.

\item Our machine learning models, built on the codebook features, are able to effectively distinguish between an image meme that will go viral and one that will not; 
  \revision{the best classifier, Random Forest, achieves 0.866 Area Under the Curve (AUC).} %
\revision{\item Our models trained on /pol/ image memes help characterizing popular memes that appear on mainstream Web communities too.
  Classifiers trained on /pol/ data correctly predict as viral 19 out of the 20 most popular image memes shared on Twitter and Reddit according to previous work~\cite{zannettou2018origins}.}
\item Through a number of case studies, we show that our features can help explaining why \newrevision{certain image memes became popular while others experienced a very limited distribution}.
\end{enumerate}

Our work sheds light on the visual elements that affect the chances of image memes to go viral and can pave the way to additional research.
For instance, one could use them to develop more effective messages when running online information campaigns (e.g., in the health context).
Additionally, our models can be relied upon to factor potential virality into moderation efforts by online social networks, e.g., by prioritizing viral imagery which could harm a larger audience, \newrevision{while discarding images that are unlikely to be re-shared}. 

\descr{Disclaimer.} Images studied in this paper come from 4chan's Politically Incorrect Board (/pol/), which is a community known for posting hateful content.
Throughout the rest of the paper, we do not censor any content or images, therefore we warn that some readers might find them disturbing.

\descr{Paper Organization.} \revision{The rest of the paper is organized as follows.
Next section introduces background and related work, and provide our definition of image memes used in this paper, while in Section~\ref{sec:dataset}, we describe our dataset. 
In Section~\ref{sec:framework}, we discuss our three research hypotheses regarding image's composition, subjects, audience, which inform the development of a codebook based on 4chan.
Next, we discuss the annotation process in Section~\ref{sec:annotation}, while in Section~\ref{sec:ml} we show that the features from our codebook can be used to train machine learning to predict virality and also 
generalize on Twitter and Reddit.
In Section~\ref{sec:case-study}, we discuss several case studies of viral and non-viral memes across platforms;
finally, we discuss the implications of our results, as well as future work, and conclude the paper in Section~\ref{sec:discussion}.
}

\section{Background and Related Work}
\label{sec:background}

In this section, we discuss %
background information, including providing the operational definition of \emph{image memes} that we use in the rest of the paper.
Then, we review relevant prior work.

\begin{figure*}  
\centering
\begin{subfigure}[b]{.3\textwidth}
\includegraphics[width=0.9\linewidth]{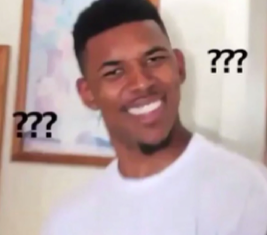}
\end{subfigure}%
~~
\begin{subfigure}[b]{.3\textwidth}
\includegraphics[width=0.9\linewidth]{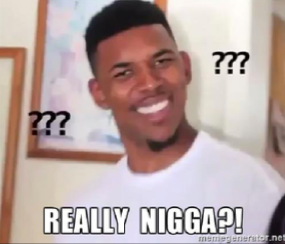}
\end{subfigure}
~~
\begin{subfigure}[b]{.3\textwidth}
\includegraphics[width=0.9\linewidth]{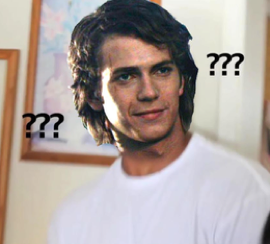}
\end{subfigure}
\caption{Viral ``Nick Young'' series meets all the conditions of a meme in this paper}
\label{fig/viralnickyoung}
\end{figure*} 

\subsection{Memes}

The term ``meme'' was coined by Richard Dawkins in 1976 to designate ``any non-genetic behavior, going through the population by variation, selection, and retention, competing for the attention of hosts''~\cite{dawkins1976selfish}.
Two decades later, a digital version of memes came into the sight of academics. 
Sociologists first noticed a ``new genre of cut and paste'' jokes in response to the 9/11 terrorist attacks, which were ``parodying, mimicking, and recycling content, embedding it in visual media culture''~\cite{kuipers2015}. 
This creation of new visual genres of expression consisted of various types of content, like catchphrases, images, and video clips, with multiple people sharing and modifying them~\cite{knobel2007online}.
Overall, memes can be defined as a piece of culture, typically with sarcastic or amusing undertones, which gains influence by being shared online~\cite{davison2012language}. 

\descr{Virality.} Recent research showed that image memes have become a prominent way for social media users to communicate complex concepts~\cite{zannettou2018origins,dupuis2019spread}.
Consistent with previous meme studies~\cite{weng2013virality}, we use the number of times a meme is shared on social media as an indicator of its virality; viral memes appear at a high frequency on online services and are shared, transformed, and imitated by many people.

\descr{Image memes.} %
In the rest of the paper, we use the following operational definition.
To be considered as an {\em image meme}, an image has to meet the two following conditions:
\begin{enumerate}
\item It must be shared by more than one user on social media;
\item It must present at least one variation on social media.
\end{enumerate}
\noindent Figure~\ref{fig/viralnickyoung} provides an example of a set of images that satisfy this definition; the example shows the ``Nick Young'' meme, which presents three variations. %

\subsection{Why and How Do People %
Share Memes?}
Users are not a passive audience of media~\cite{bryant2009media}. %
The emerging phenomenon of using memes in online communication is supported by the uses and gratification theory (UGTheory)~\cite{ruggiero2000uses}.  
The UGTheory, extended from Maslow's hierarchy of needs pyramid~\cite{maslow1943theory}, postulates that the consumers of media choose what content to consume to meet their needs and achieve gratification. 
In this context, memes may function as a tool to connect users and their audience, and be used to share information and emotions~\cite{leach2017social,sprecher2013taking}.

Moreover, different users who see a meme can behave differently, by either passively consuming it without understanding it, or by getting actively involved in the dissemination process by re-sharing or even by modifying it. 
Research in psychology showed that different personality traits, such as narcissism~\cite{leung2013generational}, extraversion~\cite{mottram2009extraversion}, and anxiety~\cite{indian2014facebook} affect the way in which people use social media, particularly with respect to how they perceive and react to online communication.

Previous research %
has also highlighted the role of ``influencers'' in the early spreading stage within online communities~\cite{weng2013virality}. %
More specifically, high homogeneity within online communities and the reputation of influencers induces conformity behavior~\cite{cialdini2004social} in users, which helps their content go viral. 
A matching attitude, beliefs, and behaviors towards a viral meme can lead users to share and imitate it under unconscious conformity to the community~\cite{cialdini2004social}. 
Additionally, research showed that individuals prefer images, or both visual and verbal information in communication~\cite{sojka2006communicating}.
Furthermore, the ``pictorial superiority effect''~\cite{nelson1976pictorial} suggests that between an image and a textual label describing it, the image is more likely to catch the attention of the user. 

\subsection{Related Work}\label{sec:related}

\descr{Studies on viral textual content.}
Previous research studied how textual information goes viral on social media, and in particular which elements influence the virality of this content.
A variety of features were tested to distinguish textual content that goes viral from the one that does not, including comments~\cite{lerman2010using}, votes~\cite{jamali2009digging}, and user-defined groups~\cite{suh2010want,yang2012we}.
Based on these features, supervised learning can be used to predict whether or not textual memes will go viral.

When studying indicators of the success of textual content, previous research found that the success of online content depends on timing, social network structure of the account posting it, randomness, and many other factors~\cite{centola2010spread,weng2012competition}. 
Other research found that textual content might become viral simply because it appeals to its audience~\cite{berger2012makes,cataldi2010emerging}. 
Another line of work argued that the reason for virality lies on whether the content generates emotions (surprise, interest, or even anxiety/anger).
However, given the competitive attention capturing environment of social network platforms, innate appeal alone may not be able to paint the whole picture of why textual memes go viral~\cite{salganik2006experimental,kitsak2010identification,yang2012we}.
Finally, some researchers argue that virality is just a random event~\cite{cashmore2009youtube}.

\descr{Studies on image memes.}
Researchers recently started studying image memes and how online users interact with them. 
Dupuis et al.~performed a survey to identify personality traits of users that are more likely to share image memes containing disinformation~\cite{dupuis2019spread}.
Crovitz and Moran provided a qualitative analysis of image memes used as a vehicle of disinformation on social media~\cite{crovitz2020analyzing}.
Zannettou et al.~\cite{zannettou2018origins} presented a large scale quantitative measurement of image meme dissemination on the Web, finding that small polarized communities like 4chan's /pol/ and The\_Donald subreddit are particularly effective in pushing racist and hateful content on mainstream social media like Twitter and Reddit.

\descr{Studies on viral marketing.}
\revision{
Previous studies on viral marketing found that targeting platforms, narrative advertising, and eliciting high arousal emotion promote sharing behavior.
Also, virality can be elicited by placing advertisements on those Web communities where the target audience is most likely to view and share them~\cite{nelson2013more}.
Online advertisements that trigger high arousal emotions, both positive and negative, such as anger, anxiety, exhilaration, amusement, are more likely to be shared than those that elicit low arousal emotions, such as sadness or contentment~\cite{berger2011arousal,berger2012makes,nelson2013emotions}.} 
\revision{
Prior work in marketing also helps explain why people share online content.
For instance, research suggests that affiliation is the human need to belong and form relationships~\cite{baumeister1995need}. 
Online users may therefore share different forms of content on social media to invite connection and interaction with others~\cite{ellison2007benefits, hayes2014social}.}

\descr{\em Novelty:} Overall, our work is, to the best of our knowledge, the first to investigate whether the visual features of an image meme contribute to its virality.

\section{Dataset}
\label{sec:dataset}

To investigate the three research hypotheses highlighted in Section~\ref{sec:introduction}, we collect examples of image memes shared on 4chan's Politically Incorrect Board (/pol/).
We choose to focus on this community because previous work showed that its users are particularly successful in influencing Internet culture and in creating image memes that will later go viral~\cite{hine2017kek,nagle2017kill,zannettou2018origins}.
In Section~\ref{sec:case-study}, we will show that, although we only use an arguably small Web community to develop our codebook and identify indicators of virality, these indicators are in fact helpful to characterize the most viral image memes that were shared on large mainstream Web communities like Twitter and Reddit.

We use a dataset of image memes collected by Zannettou et al.~\cite{zannettou2018origins}. %
The authors collected a set of 160M images from Twitter, Reddit, 4chan, and Gab, and developed a processing pipeline that clusters together visually similar images. %
Specifically, they performed clustering on all images posted on 4chan's /pol/, The\_Donald subreddit, and Gab over the course of 13 months between July 2016 and July 2017.
Using ground truth data from Know Your Meme~\cite{ref76}, a comprehensive encyclopedia of memes, they annotated each cluster by comparing the ground truth data with the cluster's medoid.\footnote{The medoid is the point in the cluster with the minimum average distance from all points in the cluster.}
Finally, they mapped all images posted on Twitter, Reddit, 4chan, and Gab, to the annotated clusters, hence obtaining all posts from these four Web communities that include image memes.

We obtain the 38,851 clusters that contain image memes posted on 4chan's /pol/ from the authors of~\cite{zannettou2018origins}.
In total, these memes appeared in 1.3M posts on /pol/.
We use the number of posts that contain a certain image meme as an indicator of its virality, i.e., we rank all image memes by the number of /pol/ posts that contain them, and consider memes at the top of the list as viral and those at the bottom as non-viral.
Following this method, we extract two datasets:
\begin{enumerate}
 \item A sample of 100 viral and 100 non-viral clusters; specifically, the top 100 and the bottom 100, respectively. We use this dataset to identify potential indicators of virality and build our codebook (see Section~\ref{sec:framework}).
 \item A sample of 50 viral and 50 non-viral clusters; specifically, at random from the top 1,000 and bottom 1,000 clusters, respectively.
 This dataset will be labeled by human annotators according to the codebook (see Section~\ref{sec:annotation}) and  used to train machine learning classifiers to test our research hypotheses (see Section~\ref{sec:ml}).
\end{enumerate}
For each cluster, we extract the medoid as the representative image meme of that cluster, thus allowing us to work on single images rather than clusters.

\newrevision{Note that virality is a spectrum, with some image memes shared millions of times on social media, others only shared once or twice, and most falling somewhere in between.
Since this work is the first characterization of indicators in virality of image memes, we believe that our choice of studying the most popular image memes on /pol/ and comparing them to images that did not get traction is appropriate to identify indicators that can tell the two classes apart. 
However, this is not free from limitations, which we discuss in Section~\ref{sec:limitations}.}

\section{A Codebook to Characterize Viral Memes}
\label{sec:framework}

In this section, we present our codebook, which guides the thematic annotation process for characterizing viral and not-viral image memes.

We break the development of this codebook in two main phases.
First, one of the authors manually analyzed the dataset of \newrevision{ the 100 most viral and 100 least viral image memes described above and came up with potential indicators of virality for the dataset.} %
\revision{Next, all six authors further characterized the image memes in the dataset with the indicators identified in the first step to produce initial codes for further annotation, using thematic coding~\cite{braun2006using}.}

\revision{
More precisely, we followed these three steps:
\begin{enumerate}
\item We discussed these initial codes and went through multiple iterations, using a portion of the data to build a final codebook. 
The process continued until the codebook reached stability and additional iterations would not refine it further.
\item To investigate the common agreement on the codebook by multiple annotators, we had them rate a portion of our dataset and discuss disagreements until a good agreement is reached.
\item We annotated the rest of our dataset and calculated a final agreement.
\end{enumerate}
}

\begin{figure*}[t]  
\centering
\begin{subfigure}[b]{.3\textwidth}
\includegraphics[width=0.9\linewidth]{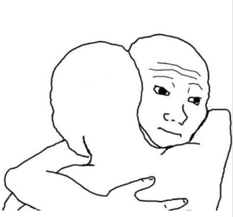}
\caption{I know that feel bro}\label{fig/f1a}
\end{subfigure}%
~~
\begin{subfigure}[b]{.5\textwidth}
\includegraphics[width=0.9\linewidth]{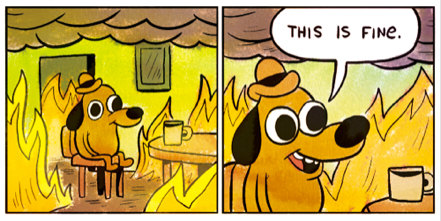}
\caption{This is fine}\label{fig/f1b}
\end{subfigure}
\caption{Example of number of panels: (a) single and (b) multiple.}
\label{fig/f1}
\end{figure*}

\revision{Drawing from research in a number of fields, we identify a number of elements (\emph{``features''}) that are potentially characteristic of image meme virality and that can help us answer our three research hypotheses.}
In the rest of this section, we describe these features in detail, along with the motivation for selecting them, grouping them in three sections based on which research hypotheses they help answering: \emph{composition} (RH1), \emph{subjects} (RH2), or \emph{audience} (RH3).

\subsection{RH1: Composition}

Our first research hypothesis is that the composition of an image contributes to making an image meme go viral. 
In visual arts, composition refers to the organization of visual elements in a picture, including color, form, line, shape, space, texture, and value. 
The different approaches to good composition obey the principles of the arts, which take into account the balance of an image, its emphasis, movement, proportion, rhythm, unity, and variety~\cite{collingwood1958principles}.
\newrevision{While composition might not be the only factor determining if an image meme will go viral (e.g., a well-designed image may not succeed as a meme because users will fail to understand its meaning), we hypothesize that a poorly composed image is unlikely to become viral in the first place.}

Previous research found that when viewing a scene, people focus on its center first, in what is known as \emph{center bias}~\cite{buswell1935people,parkhurst2003scene,tatler2007central}, and that the way in which objects are arranged in an image affects whether viewers perceive them as salient or not~\cite{itti2001computational}.
Additionally, researchers found that movement in a scene is able to capture viewers' attention~\cite{johansson1975visual}.

Based on previous research on image composition and manual review of the 100 viral and non-viral memes from /pol/, we decided to use four features to investigate RH1: the number of panels in it, its type, its scale, and the type of movement in it.
In the following, we describe these four features in detail.

\descr{F.1 Number of panels.} %
By analyzing our dataset of 200 memes, we found that image memes may be composed of a single panel or of multiple ones, similar to comic book strips. %
We argue that multiple panels take longer time to read than a single image, therefore this might have an impact in gaining viewers' attention.
Therefore, we use the number of panels in a meme to understand whether this element can affect its virality.
In particular, we distinguish the following two cases:

\begin{itemize}
  \item Single panel: memes that are composed of only one image; e.g., see Figure \ref{fig/f1a}, ``I know that feel bro,'' from the famous Wojak series, which is used in expression empathy or agreement to one's expression.

  \item Multiple panels: memes that are composed of a series of images; e.g., see Figure \ref{fig/f1b}, ``This is fine,'' originally coming from the comic series of K.C. Green's Gunshow, comic number 648 and used to convey hopeless emotion in a despair situation.
\end{itemize}

\begin{figure*}[t]
\centering
\begin{subfigure}[b]{.2\textwidth}
\includegraphics[width=0.9\linewidth]{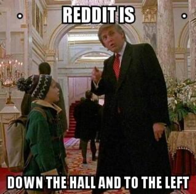}
\caption{}\label{fig/f2a}
\end{subfigure}%
~~
\begin{subfigure}[b]{.38\textwidth}
\includegraphics[width=0.9\linewidth]{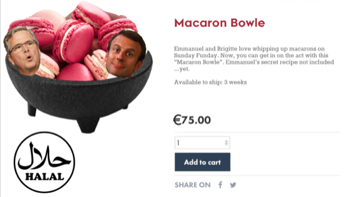}
\caption{}\label{fig/f2b}
\end{subfigure}
~~
\begin{subfigure}[b]{.3\textwidth}
\includegraphics[width=0.9\linewidth]{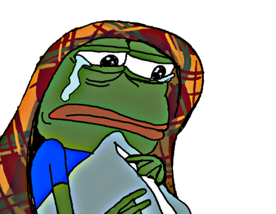}
\caption{}\label{fig/f2c}
\end{subfigure}
\caption{Type of images: (a) photo, (b) screenshot, and (c) illustration}
\label{fig/f2}
\end{figure*}

\descr{F.2 Type of the images.} %
Image memes do not only come in the form of illustrations, but also use photographs or screenshots, as confirmed by the sample dataset that we examined to build our codebook (see Section~\ref{sec:dataset} and Figures~\ref{fig/f2a},~\ref{fig/f2b}, and~\ref{fig/f2c} for examples).
We are therefore interested in understanding whether certain types of images better capture viewers' attention.
To this end, we consider three types of images:
\begin{itemize}
  \item Photo: a picture taken by a camera (e.g., Figure~\ref{fig/f2a}).
  \item Screenshot: an image of a screenshot taken from a computer screen, for example of part of a Web page (e.g., Figure~\ref{fig/f2b}). 
  \item Illustration: a drawing, painting, or printed work of art (e.g., Figure \ref{fig/f2c}).
\end{itemize}

\begin{figure*}[t]
\centering
\begin{subfigure}[b]{.25\textwidth}
\includegraphics[width=0.9\linewidth]{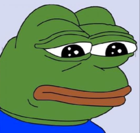}
\caption{}\label{fig/f4a}
\end{subfigure}%
~~
\begin{subfigure}[b]{.23\textwidth}
\includegraphics[width=0.9\linewidth]{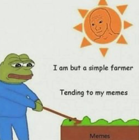}
\caption{}\label{fig/f4b}
\end{subfigure}
~~
\begin{subfigure}[b]{.3\textwidth}
\includegraphics[width=0.9\linewidth]{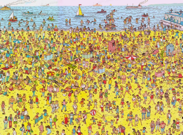}
\caption{}\label{fig/f4c}
\end{subfigure}
\caption{Scale: (a) close up, (b) medium shot, and (c) long shot.}
\label{fig/f4}
\end{figure*}

\descr{F.3 Scale.} %
Research in human vision showed that viewers' gaze is biased towards the center of a scene (i.e., center bias)~\cite{buswell1935people,parkhurst2003scene,tatler2007central}. %
We hypothesize that an image that is a close up of a subject will facilitate viewer focus on the salient part of the image and therefore catch their attention, while a large scale scene in which it is hard to identify the part to focus on might fail in attracting the viewer's attention.
To investigate how these aspects might affect the virality of an image meme, we consider its scale, which takes into account how the main subject is put in relation with the layout of the remaining elements of the image. %
Based on the definitions of shots used in film studies~\cite{arijon1991grammar,tsingalis2012svm}, we define three scales for images:
\begin{itemize}
  \item Close up: a shot that tightly frames a person or object, such as the subject's face taking up the whole frame (e.g., Figure~\ref{fig/f4a}).
  \item Medium shot: a shot that shows equality between subjects and background, such as when the shot is ``cutting the person in half'' (e.g., Figure~\ref{fig/f4b}).
  \item Long shot: a shot where the subject is no longer identifiable and the focus is on the larges scene rather than on one subject (e.g., Figure~\ref{fig/f4c}).
\end{itemize}

\descr{F.4 Movement.} Research in psychonomics found that when watching a scene, movement is able to effectively capture the viewer's attention~\cite{johansson1975visual}.
Artists have been following a similar intuition for centuries, and have used a variety of techniques to provide the illusion of movement in their paintings~\cite{gottlieb1958movement}. %
In this paper, we hypothesize that the perception of movement in an image meme might contribute to catching the viewer's attention and therefore influence its chances to go viral.

To characterize the type of movement in an image, we identify three types: physical movement, emotional movement, and causal movement.
A meme might contain different types of movement at the same time. %
In our codebook, we consider whether images indicate movement (as identified by arts research~\cite{gottlieb1958movement}), otherwise we consider them as presenting ``no movement.''

\begin{figure*}[t]
\centering
\begin{subfigure}[b]{.28\textwidth}
\includegraphics[width=0.9\linewidth]{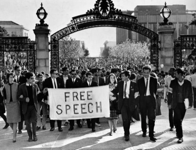}
\caption{}\label{fig/f5a}
\end{subfigure}%
~~
\begin{subfigure}[b]{.28\textwidth}
\includegraphics[width=0.9\linewidth]{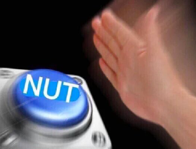}
\caption{}\label{fig/f5b}
\end{subfigure}
~~
\begin{subfigure}[b]{.38\textwidth}
\includegraphics[width=0.9\linewidth]{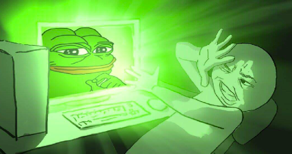}
\caption{}\label{fig/f5c}
\end{subfigure}
\caption{Examples of movement: (a) an image that includes physical movement, (b) an image that combines physical movement and causal movement, (c) an image that contains physical movement, emotional movement, and causal movement.}
\label{fig/f5}
\end{figure*}

We classify all movement in images as physical movement, while an emotional expression on the face or in the body language is also annotated as emotional movement.
We categorize movement as causal when the movement sequence is caused by one component (sender) to another (recipient). 
Taking Figure~\ref{fig/f5b} as an example, when an initially stationary object, (the button in Figure~\ref{fig/f5b}) is being set into motion by another moving object (the hand), viewers spontaneously interpret the sequence as being causal~\cite{kerzel2000launching}.

Figure~\ref{fig/f5} contains examples of the types of movement that we consider: (a) a meme that contains physical movement only (people walking); (b) a meme that contains physical movement (the movement of the hand) and causal movement (the movement of the hand causes the movement of the button, i.e., the button acts as a recipient) (c) a meme that contains physical movement (the character turning away from the screen), emotional movement (the character showing a frightened/protective facial expression and body language with the action of moving backwards), and causal movement (the sudden change of the screen causes the response of the character).

\begin{figure*}[t]
\centering
\begin{subfigure}[b]{.25\textwidth}
\includegraphics[width=0.9\linewidth]{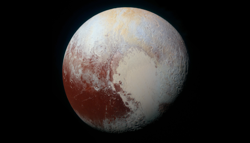}
\caption{}\label{fig/f6a}
\end{subfigure}%
~~
\begin{subfigure}[b]{.25\textwidth}
\includegraphics[width=0.9\linewidth]{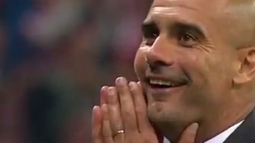}
\caption{}\label{fig/f6b}
\end{subfigure}
~~
\begin{subfigure}[b]{.2\textwidth}
\includegraphics[width=0.9\linewidth]{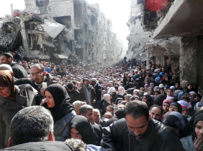}
\caption{}\label{fig/f6c}
\end{subfigure}
~~
\begin{subfigure}[b]{.18\textwidth}
\includegraphics[width=0.9\linewidth]{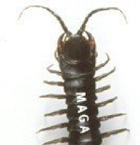}
\caption{}\label{fig/f6d}
\end{subfigure}
\caption{Subject of memes: (a) object, (b) character, (c) scene, (d) creature.}
\label{fig/f6}
\end{figure*}

\subsection{RH2: Subjects}

Our second research hypothesis is that the subjects depicted in an image meme have an effect on whether the meme goes viral or not.
In visual arts, the subject of an image refers to the person, object, scene, or event described or represented in it~\cite{shatford1986analyzing}.
Basically, the subject is the essence (main idea) of the work. 
\newrevision{We hypothesize that an image that does not have a clear subject that the viewer can focus their attention on is unlikely to go viral.}
To study this hypothesis, we look at the type of subject (e.g., whether the focus of the image is on a character or an object) as well as at the characteristics of this subject (e.g., their facial expression).

\descr{F.5 Type of subject.} %
The types of subjects that can be depicted in an image include landscapes, still life, animals, and portraits of people~\cite{shatford1986analyzing}.
Research in human vision found that viewers' attention tends to be attracted by the faces of the characters in a picture~\cite{buswell1935people}.
Looking at the 200 memes in our dataset, we identify four types of subjects appearing in them: characters, scenes, creatures, and objects (e.g., see Figure~\ref{fig/f6}). 
Images that do not contain any of these types of subjects are categorized as ``other.''
More precisely, we characterize subjects as follows: %
\begin{itemize}
  \item ``Object'' refers to a material thing that can be seen and touched, like a table, a bottle, a building, or even a celestial body (see Figure~\ref{fig/f6a}). %
  \item ``Character'' refers to people (see Figure~\ref{fig/f6b}) or anthropomorphized creatures/objects, such as cartoon characters. %
  \item We categorize a subject as ``scene'' when the situation or activity depicted in an image meme is its main focus, instead of it being on the single characters or objects depicted in it (see Figure~\ref{fig/f6c}). %
  \item ``Creature'' refers to an animal that is not anthropomorphized (see Figure~\ref{fig/f6d})~\cite{opfer2011development}. %
\end{itemize}

\begin{figure*}[t]
\centering
\begin{subfigure}[b]{.13\textwidth}
\includegraphics[width=0.9\linewidth]{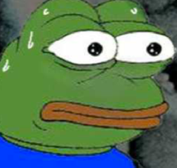}
\caption{}\label{fig/f7a}
\end{subfigure}%
~~
\begin{subfigure}[b]{.13\textwidth}
\includegraphics[width=0.9\linewidth]{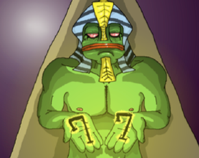}
\caption{}\label{fig/f7b}
\end{subfigure}
~~
\begin{subfigure}[b]{.13\textwidth}
\includegraphics[width=0.9\linewidth]{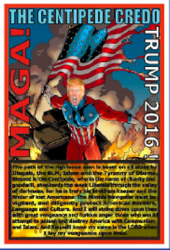}
\caption{}\label{fig/f7c}
\end{subfigure}
~~
\begin{subfigure}[b]{.13\textwidth}
\includegraphics[width=0.9\linewidth]{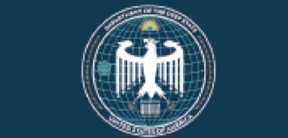}
\caption{}\label{fig/f7d}
\end{subfigure}
~~
\begin{subfigure}[b]{.13\textwidth}
\includegraphics[width=0.9\linewidth]{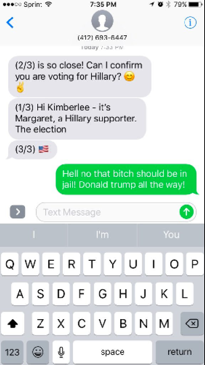}
\caption{}\label{fig/f7e}
\end{subfigure}
~~
\begin{subfigure}[b]{.13\textwidth}
\includegraphics[width=0.9\linewidth]{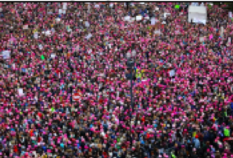}
\caption{}\label{fig/f7f}
\end{subfigure}
~~
\begin{subfigure}[b]{.13\textwidth}
\includegraphics[width=0.9\linewidth]{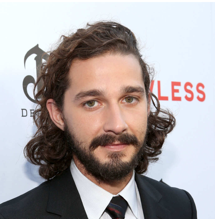}
\caption{}\label{fig/f7g}
\end{subfigure}
\caption{Attributes of the subject: (a) Facial expression, (b) posture, (c) poster, (d) sign, (e) screenshot, (f) scene, and (g) unprocessed photo.}
\label{fig/f7}
\end{figure*}

\descr{F.6 Attributes of the subject.} %
For each category, we provide subcategories to further analyze the attributes of the subjects in the meme.
For images whose subject is one or more characters, we consider whether the image's visual attraction lies with the character's facial expression (e.g., Figure~\ref{fig/f7a}) or with their posture (e.g., Figure~\ref{fig/f7b}).
In the next section (F.7), we will describe how we further refine the character's emotion. 
For the other attributes, we identify five features:

\begin{itemize}
  \item Poster Figure~\ref{fig/f7c}: informative large scale image including both textual and graphic elements.
  There are also posters only with either of these two elements. 
  Posters are generally designed to be displayed at a long-distance.
  \item Sign Figure~\ref{fig/f7d}: informs or instructs the viewer through text, symbols, graph, or a combination of these.
  \item Screenshot Figure~\ref{fig/f7e}: is a digital image that shows the contents of a electronic screen display.
  \item Scene Figure~\ref{fig/f7f}: a place where an event occurs.
  \item Unprocessed photo Figure~\ref{fig/f7g}: raw photo taken by a camera without being modified.
\end{itemize}

\begin{figure*}[t]
\centering
\begin{subfigure}[b]{.225\textwidth}
\includegraphics[width=0.9\linewidth]{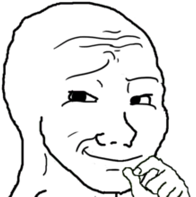}
\caption{}\label{fig/f8a}
\end{subfigure}%
~~
\begin{subfigure}[b]{.2\textwidth}
\includegraphics[width=0.9\linewidth]{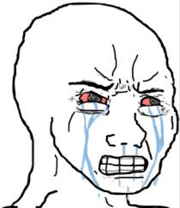}
\caption{}\label{fig/f8b}
\end{subfigure}
~~
\begin{subfigure}[b]{.225\textwidth}
\includegraphics[width=0.9\linewidth]{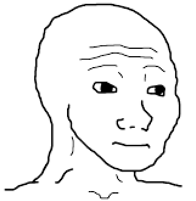}
\caption{}\label{fig/f8c}
\end{subfigure}
\caption{Emotion: (a) positive, (b) negative, and (c) neutral.}
\label{fig/f8}
\end{figure*}

\descr{F.7 Character's emotion.} %
Research showed that the emotion perceived from images causes physiological reactions in viewers~\cite{hillman2004emotion} %
and can even speed up the perception of time~\cite{lui2011emotion}. %
Research in neuroscience showed that faces that exhibit an emotion capture people's attention~\cite{vuilleumier2002facial}. %
Based on this research, we hypothesize that the emotions portrayed by characters in an image meme might have an impact on capturing viewers' attention and in their decision to re-share the meme.
To study this, we include a character's emotion as one of the features in our study. 
We consider images that have been previously annotated as ``facial expression'' or ``posture'' in F6 to be further refined, and define three states of emotions: positive, negative, and neutral (See Figure~\ref{fig/f8}).  
All the emotions are annotated based on the character's facial expression or body language.
Strong indications of emotions is annotated as positive (e.g., Figure~\ref{fig/f8a}) or negative (e.g., Figure~\ref{fig/f8b}). 
Otherwise, we consider the emotion of the character as neutral (see Figure~\ref{fig/f8c}).  
Typical positive emotions reflected in image memes are: laughing, smiling, smug, excited, while typical negative emotions reflected in image memes are: crying, being angry/nervous, showing impatience/boredom/shyness. 

\begin{figure}[t]  
\centering
\begin{subfigure}[b]{.25\textwidth}
\includegraphics[width=0.9\linewidth]{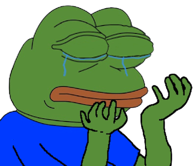}
\caption{}\label{fig/f9a}
\end{subfigure}%
\hspace*{-0.2cm}
\begin{subfigure}[b]{.25\textwidth}
\includegraphics[width=0.9\linewidth]{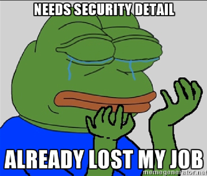}
\caption{}\label{fig/f9b}
\end{subfigure}
\caption{Examples of image memes with: (a) no words, and (b) with words.}
\label{fig/f9}
\end{figure}

\descr{F.8 Contains words.} %
Image memes frequently include a word caption to better elicit the message of the meme (see Figure~\ref{fig/f9}).
However, lengthy text can delay people's recognition of a meme~\cite{van2009split}. %
We therefore hypothesize that lengthy text potentially impairs a meme's virality.
To better study consider the number of words in a meme as a feature.

\subsection{RH3: Audience}

Our third hypothesis deals with the fact that the intended audience of an image meme can influence whether the meme goes viral or not.
We define the audience of an image meme as the set of viewers who fully understand the meme when they encounter it~\cite{borzsei2013makes}. %
The users of different sub-communities have different backgrounds and sub-cultures, and the memes that they produce might contain elements that resonate with their community's culture, but might not be easily understood by users who are not familiar with it.
To understand how limiting the intended audience of an image meme might impact its virality, we first attempt to determine if a meme is intended for a general audience (i.e., does not require any specific knowledge) or for a particular one.
Informed by previous research showing that hateful memes often go viral on social media~\cite{zannettou2018origins}, we then aim to understand if specific tones of the meme (i.e., hateful, racist, and political tones) have an effect on its virality.

\begin{figure}[t]  
\centering
\begin{subfigure}[b]{.25\textwidth}
\includegraphics[width=0.9\linewidth]{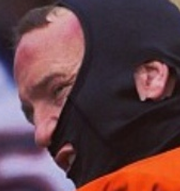}
\caption{}\label{fig/f10a}
\end{subfigure}%
~
\begin{subfigure}[b]{.25\textwidth}
\includegraphics[width=0.9\linewidth]{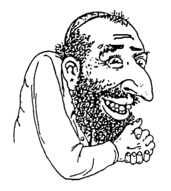}
\caption{}\label{fig/f10b}
\end{subfigure}
\caption{(a) The meme ``manning face'' as an example of ``human common'' (b) The meme ``happy merchant'' as an example of ``cultural specific'' meme}
\label{fig/f10}
\end{figure}

\descr{F.9 Intended audience.} %
We distinguish image memes into two categories according to their intended audience: human common and culture specific. 
Human common refers to those image memes that arouse the common experiences and emotions of any viewer.
These are supposed to be understood by all social media users regardless of their background (e.g., Figure~\ref{fig/f10a}).
Marketing research suggests that visual media that uses everyday scenes and familiar situations is more likely to resonate with a general audience~\cite{stutts1999use}. %
In the context of virality, we hypothesize that these memes are more likely to go viral on social media.
Culture specific memes are those that require some background knowledge to be fully understood (e.g., Figure~\ref{fig/f10b}); in this case, only members of the intended community can understand the full meaning of the meme, and this can affect its virality.

Based on prior work suggesting that political and hateful memes are particularly likely to be re-shared, especially by polarized communities~\cite{zannettou2018origins}, we further categorize culture specific memes into hateful, racist, and political. 
Previous research argued that polarized communities like 4chan are working on ``attention hacking,'' which is a way to propagate their idea by sharing viral memes~\cite{marwick2017media}.

\section{Annotation}
\label{sec:annotation}

We now discuss the methodology followed to annotate our dataset of 100 image memes (50 viral and 50 non-viral; see Section~\ref{sec:dataset}) using the codebook presented above.

\descr{Human Annotators.} The process of labeling images cannot be entirely automated for two reasons.
First, while computer vision has made tremendous progress in automatically recognizing objects in images~\cite{karpathy2015deep,vinyals2016show}, image memes often are not well-polished pictures; rather, they include a mixture of drawings and collages.
Second, most of the features that we identified are highly subjective (e.g., human common vs cultural specific), and it would be difficult for an automated approach to label them correctly.  
Hence, we opt to rely on human annotators.
To this end, we developed a Web interface to facilitate annotation, and had six annotators use it to label the 100 image memes.
Note, however, that we do automate the extraction of words from the image (see F.8 in Section~\ref{sec:framework}), using optical character recognition (OCR) techniques.

\subsection{Annotation Platform}

The majority of our codebook includes non-binary choices and has branching depending on some of the choices selected.
To address this, we built a custom codebook-oriented annotation platform, delivered as a Web application.
The structure of a codebook is encoded as a graph, which in turn is stored as JSON in a database.
Annotation questions are linked to nodes in this graph, which allows us to capture the hierarchical dimensions of the codebook, as well as provide direct reference points into the codebook for annotators.

It is important to note that some of our features have mutually exclusive labels and others do not.
Since annotators can choose multiple labels for these features, in the end, we have 35 possible labels across 9 features for each image we annotate.
Table~\ref{tab:fleiss} lists the 35 labels that we used for annotation.

\subsection{Annotation Process}

Six annotators used the Web application presented above, i.e., they were shown images from both viral and non-viral clusters and asked to label based on the codebook presented in Section~\ref{sec:framework}.
Note that three annotators were male and three females, and all of them were in their 20s and 30s and had a graduate degree. 
\newrevision{The annotators have extensive experience with memes in general and hateful content, which makes it easier to understand certain coded messages that might appear in a meme.}

\descr{Ethical and Privacy Considerations.} Our study was approved by the IRB at Boston University.
More specifically, since participants were only able to choose from a set of pre-determined options and no sensitive information was collected in the process, the IRB granted us an exemption.

\revision{Note that human faces appear in our dataset, which might prompt privacy concerns. 
However, all images with human faces included in this paper belong to public figures (e.g., celebrities), therefore, we argue that the privacy implications of including their faces are minimal.
For example, Nick Young is a famous basketball player and Peyton Manning was a star NFL Quarterback. 
All other figures in the paper are either drawings or are large-scale scenes where the focus is not on the individuals.
Regardless, we further discuss potential privacy issues in Section~\ref{sec:discussion}.}

\begin{table*}
\small
\centering
  \begin{tabular}{
     >{\raggedright\arraybackslash}p{0.13\textwidth}@{}rl|>{\raggedright\arraybackslash}p{0.13\textwidth}@{}rl}
    \toprule
     \bf{Feature}&\bf{Label} & \bf{Fleiss} & \bf{Feature} &  \bf{Label} & \bf{Fleiss} \\
    \midrule
    F.1 Number of panels & & & F.6 Attributes of subject & \\
    & A single panel & 0.958 *** && Facial expression & 0.709**\\
    & Multiple panels & 0.958 *** && Stationary pose/posture & 0.649 **\\
    F.2 Image type &&&& Poster & 0.674  **\\
    & Photo & 0.926 *** && Sign & 0.640  ** \\
    & Illustration & 0.947   *** && Screenshot & 0.933  ***\\
    & Screenshot & 0.951   *** && Situation & 0.674  **\\
    & None of the above & 1.000   *** && Unprocessed photo & 0.772  **\\
    F.3 Scale &&&&Other	& 0.624  **\\
    & Close up & 0.666 ** & F.7 Emotion \\
    & Medium shot & 0.813 *** && Positive & 0.794  **\\
    & Long shot & 0.895  *** && Negative & 0.765  **\\
    F.4 Movement &&&& Neutral & 0.627  **\\
    & Physical Movement & 0.724  ** & F.9 Audience\\
    & Emotional Movement & 0.125 && Human Common	& 0.719  ** \\
    & Causal Movement & 0.509  * && Cultural Specific & 0.729  **  \\
    & No movement & 0.366 && Hateful & 0.333 \\
    F.5 Type \hspace*{-0.1cm} of subject &&&& Political & 0.876  ***\\
    & An object/objects & 0.592  * && Racist & 0.805  ***\\
    & A character/characters & 0.768  ** && None of above & 0.837  ***  \\
    & A scene/scenes & 0.544  * &&\\
    & A creature/creatures & 0.457  * &&\\
    & None of above	& 1.000  ***  &&\\
    \bottomrule
\end{tabular}
  \caption{Inter-annotators agreement for each feature~\cite{landis1977measurement}. *** indicates almost perfect agreement (Fleiss score above 0.8), ** substantial agreement (Fleiss score between 0.6 and 0.8), while * indicates moderate agreement (Fleiss score between 0.4 and 0.6).} %
  \label{tab:fleiss}
  \vspace{-0.4cm}
\end{table*}

\descr{Inter-annotator Agreement.}
Next, we measure the agreement among annotators using Fleiss' Kappa score~\cite{landis1977measurement}.
We do so aiming to understand if our codebook features are intuitive and humans can reliably identify them in images, as well as to establish whether the labeling is ``reliable'' enough.
Note that the Fleiss' Kappa score ranges from 0 to 1, where 0 indicates no agreement and 1 perfect agreement.
To determine if annotators can identify the different aspects of each feature, we break them into the corresponding possible values; for instance, the movement feature has five labels: physical, causal, emotional, no movement, and none of above.

In Table~\ref{tab:fleiss}, we report the Fleiss' Kappa scores for each feature/option.
Overall, the agreement between annotators is generally high: 13 of the 35 labels have  almost perfect agreement (score above 0.8), 15 substantial agreement (score between 0.6 and 0.8), and 4 moderate (score between 0.4 and 0.6).
Only 3 labels fall below the threshold of moderate agreement, specifically, ``Emotional movement,'' ``No movement'' and ``Hateful.''
For ``Emotional Movement'' and ``No Movement,'' we believe that the agreement is low because, although our codebook has a definition for it, the perception of emotion is subjective, and different annotators can perceive the same image as emotional or not.
Thus, ``Emotional Movement'' has the lowest agreement among all options.
For the hateful option, different cultural backgrounds in the annotators may result in different understandings of what constitutes a hateful meme.

\descr{Number of Words.}
As discussed in Section~\ref{sec:framework}, our codebook takes into account the presence of words in an image meme.
In particular, we hypothesize that image memes containing too many words might be less likely to go viral.
Unlike other features, which need to be manually annotated, we can automatically extract the words in an image meme using OCR techniques; specifically, we use the Optical Character Recognition (OCR) API from Google vision~\cite{mulfari2016using}.

\begin{figure}[t]
    \centering
    \includegraphics[width=0.85\linewidth]{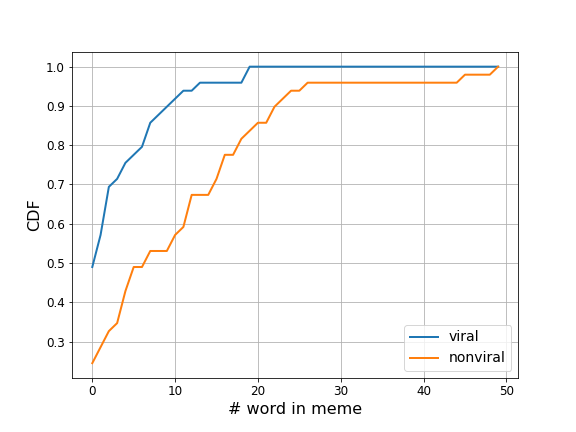}
    \caption{Distribution of the number of words in viral and non-viral memes.}
    \label{fig11}
\end{figure}

Having extracted the distribution of words across the viral and non-viral memes, we set to determine a threshold of the number of words that is likely to impair the virality of a meme. 
Later on, we will encode this threshold into a feature used by a machine learning classifier.
In Figure~\ref{fig11}, we plot the cumulative distribution functions (CDFs) of the number of words in viral and non-viral memes in our dataset.
To assess whether the difference in the distributions of the two classes is statistically significant, we run a two sample Kolmogorov Smirnoff test~\cite{massey1951kolmogorov}.
The test allows us to reject the null hypothesis that the number of words in an image meme does not have an effect on its virality ($p < 0.005$), suggesting that this is indeed a good feature to characterize viral and non-viral memes.
As we can see, the presence of words is less common in the viral class, with 50\% of the viral memes in our dataset not containing any words.
In general, the viral memes in our dataset contain less words than the non-viral ones: 94.0\% of them contain less than 15 words, while 31.91\% of the non-viral memes exceed this number.
As a result, we select 15 as the threshold of words beyond which the virality of a meme might be impaired.

\section{Modeling Indicators of Image Virality} %
\label{sec:ml}

We now investigate whether the features from our codebook can be used to train machine learning models to %
determine whether or not an image meme will go viral.
More specifically, we train several classifiers to identify which memes will go viral, and show that our features can indeed identify viral memes.
\revision{
To address the limitations of tree-based classifiers over limited training data (see Section~\ref{sec:twitterandreddit}), we report the results of both the best classifier and the best non-tree-based classifier.
The best classifier is Random Forest, with AUC 0.866, and the best non-tree-based classifier is KNN, with AUC 0.828.
}
We then discuss the importance of the features identified by our classifiers, reasoning on which elements are particularly indicative of virality/non-virality.

\subsection{Classification}

\begin{table}
\centering
\small
\setlength{\tabcolsep}{3pt}
  \begin{tabular}{l|rrr|rrr}
    \toprule
    {\bf Classifier} & AUC & std & acc & precision & recall & f1-score\\
    \midrule
    Random Forest & 0.87 & 0.13 & 0.98 & 0.98 & 0.98 & 0.98 \\
     Ada boost & 0.85 & 0.11 & 0.91 & 0.92 & 0.91 & 0.91\\
     KNN & 0.83 & 0.10 & 0.98 & 0.98 & 0.98 & 0.98\\
     SVM & 0.81 & 0.10 & 0.71 & 0.75 & 0.69 & 0.70\\
     Logistic regression & 0.81 & 0.10 & 0.77 & 0.77 & 0.77 & 0.77\\
     Gaussian Bayesian & 0.77 & 0.18 & 0.69 & 0.76 & 0.69 & 0.67\\
     Decision tree & 0.75 & 0.18 & 0.82 & 0.84 & 0.82 & 0.82\\
     Neural network & 0.79 & 0.10 & 0.90 & 0.90 & 0.90 & 0.90\\
  \bottomrule
\end{tabular}
  \caption{Performance of all classifiers.}
  \label{tab:results}
\end{table}

We select the features for our classification models as follows.
We start with the features described in Section~\ref{sec:framework}; we consider those whose options are exclusive (e.g., the scale of a meme can only be either close up, medium shot, or long shot) as a single feature,
while we split those that allow multiple options into multiple features (e.g., an image meme can present a character and an object at the same time).
This yields 30 features: %
number of images; type of images (photo, illustration, screenshot, none of above); scale of images; movement (physical movement, emotional movement, causal movement, there is no movement); content (an object/objects, a character/characters, a scene/scenes, a creature/creatures, none of above); type (facial expression, stationary pose/posture, poster, sign, screenshot, scene/situation, unprocessed photo, other); emotion; words; audience; cultural specific (hateful, political, racist, none of above).

We then select eight classifiers to train using our features: Random Forest~\cite{liaw2002classification}, Support Vector Machines (SVM)~\cite{suykens1999least}, Logistic regression~\cite{hosmer2013applied}, K-Nearest Neighbors (KNN)~\cite{weinberger2006distance}, Ada boost~\cite{hastie2009multi}, Gaussian Bayesian~\cite{williams1998bayesian}, Decision tree~\cite{safavian1991survey}, and Neural network~\cite{hecht1992theory}. 
For each classifier, we take the annotated set of 100 image memes and perform a 10-fold cross validation, i.e., randomly dividing the dataset into ten sets and using nine for training and one for testing.
We repeat this process 10 times, and calculate the Area Under the ROC Curve (AUC) as well as its standard deviation. 
The results are reported in Table~\ref{tab:results}.
\revision{
We observe that Random Forest achieves the best performance, with an AUC of 0.866.
Among non-tree based classifiers, KNN has the best classification performance, achieving an AUC of 0.828.
}

\subsection{Feature Importance}
We now set to understand which features particularly contribute to the classification decision.
To identify indicators of virality (or non-virality), we perform a feature analysis of the %
best performing classifier (Random Forest).
In the following, we discuss the top five features learned by this model, and whether images that present them are more likely to go viral or to not be shared.

\begin{enumerate}

\item \textbf{\em Facial expression:}
The Random Forest classifier picks up the facial expression of a character as the most important feature for classification.
42 out of 50 viral memes (84\%) are labeled as ``facial expression'' while 19 out of 50 non-viral ones (38\%) possess this feature.

\item \textbf{\em Character emotion:} Image memes showing a positive emotion are more likely to go viral (with 39\% of the viral memes in our dataset presenting this feature vs. 15\% of the non-viral one).
Similarly, a negative emotion contributes to virality (with 27\% of the viral memes presenting this feature compared to 17\% of the non-viral ones).

\item \textbf{\em Character posture:} 
Our model puts the posture of a character as the third most important feature. 46\% viral memes presenting this feature compared to 30\% non-viral memes.

\item \textbf{\em Image scale:} While a medium shot scale does not seem to affect virality in either direction (with 61\% and 58\% of the viral and non-viral memes in our dataset presenting this scale), we find that image memes that present a close up scale are more likely to go viral (34\% of the viral memes in our dataset present this feature, while only 14\% of the non-viral ones), while images that use a large shot scale are less likely to be viral (27\% of the non-viral memes present this feature, while only 4.6\% of the viral ones).

\item \textbf{\em Character as subject:} 
The presence of a character is the most strong indicators among other subjects in predicting virality.
93\% of the image memes in the viral class are labeled as ``character,'' compared to 67\% in the non-viral class.
\end{enumerate}

\subsection{Evaluation of the Research Hypotheses}
\label{sec:hypotheses}

We now re-evaluate the three research hypotheses after analyzing the results of our evaluation.
Our work started by setting out three research hypotheses on whether the composition (RH1), the subjects (RH2), and the audience (RH3) of an image meme have an effect on its chance of going viral.
We have then developed a codebook and a number of features to help us investigate these three hypotheses, and analyze them by training machine learning models and use them to identify the features that are important in distinguishing between viral and non-viral memes. 
Next, we summarize the main findings from our analysis and discuss whether they confirm our research hypotheses.

\descr{RH1: Composition.}
Our classification model showed that the scale of an image is highly discriminative, with images that use a close up being more likely to go viral, while those that use a long shot being more likely to not get shared.
This suggests that the composition of an image does have an effect on whether an image meme will go viral or not, confirming RH1. %

\descr{RH2: Subjects.}
Our classification model shows that image memes that contain characters are more likely to go viral and that those that contain objects are less likely to go viral.
It also finds that image memes that contain a facial expressions (both positive and negative) are more likely to go viral, as well as those images where the character has a particular posture.
This confirms that the subjects of an image do play a big role in whether the image meme will go viral, confirming RH2. %

\descr{RH3: Audience.}
We hypothesized that the target audience of an image meme influences the chances of it to go viral.
However, we do not find a confirmation of this in our classification model, since none of the audience-related features are found to be important.
Therefore, we are unable to confirm RH3.

\begin{table*}[t]
\small
\centering
\setlength{\tabcolsep}{4pt}
\begin{tabular}{@{}lrlrlrlr@{}}
\toprule
\multicolumn{2}{c}{\textbf{Reddit}}                 & \multicolumn{2}{c}{\textbf{Twitter}}                                        \\ \midrule
\multicolumn{1}{c}{\textbf{Meme}}                                              & \multicolumn{1}{c|}{\textbf{\#Posts (\%)}} & \multicolumn{1}{c}{\textbf{Meme}} & \multicolumn{1}{c}{\textbf{\#Posts(\%)}} \\ \midrule
\href{http://knowyourmeme.com/memes/manningface}{Manning Face}                       & \multicolumn{1}{r|}{12,540 (2.2\%)}       &  \href{http://knowyourmeme.com/memes/roll-safe}{Roll Safe}                                  & 55,010 (5.9\%)                                      \\
\href{http://knowyourmeme.com/memes/thats-the-joke}{That's the Joke}                    & \multicolumn{1}{r|}{7,626 (1.3\%)}         & \href{http://knowyourmeme.com/memes/evil-kermit}{Evil Kermit}                                    & 50,642 (5.4\%)                                      \\
\href{http://knowyourmeme.com/memes/feels-bad-man-sad-frog}{Feels Bad Man/ Sad Frog}           & \multicolumn{1}{r|}{7,240 (1.3\%)}      &                                 \href{http://knowyourmeme.com/memes/arthurs-fist}{Arthur's Fist} & 37,591 (4.0\%)                                      \\
\href{http://knowyourmeme.com/memes/confession-bear}{Confession Bear}                    & \multicolumn{1}{r|}{7,147 (1.3\%)}       &   \href{http://knowyourmeme.com/memes/nut-button}{Nut Button}                               &     13,598 (1,5\%)                                 \\
\href{http://knowyourmeme.com/memes/this-is-fine}{This is Fine}                           & \multicolumn{1}{r|}{5,032 (0.9\%)}        &  \href{http://knowyourmeme.com/memes/spongebob-mock}{Spongebob Mock}                                  &  11,136 (1,2\%)                                     \\
\href{http://knowyourmeme.com/memes/smug-frog}{Smug Frog}         & \multicolumn{1}{r|}{4,642 (0.8\%)}        &   \href{http://knowyourmeme.com/memes/reaction-images}{Reaction Images}                             &  9,387 (1.0\%)                                      \\
\href{http://knowyourmeme.com/memes/roll-safe}{Roll Safe}                     & \multicolumn{1}{r|}{4,523 (0.8\%)}     &      \href{http://knowyourmeme.com/memes/conceited-reaction}{Conceited Reaction}  &    9,106 (1.0\%)                                  \\
\href{http://knowyourmeme.com/memes/rage-guy-fffffuuuuuuuu}{Rage Guy}                        & \multicolumn{1}{r|}{4,491 (0.8\%)}         &  \href{http://knowyourmeme.com/memes/expanding-brain}{Expanding Brain}                                                               & 8,701 (0.9\%)                                       \\
\href{http://knowyourmeme.com/memes/make-america-great-again}{Make America Great Again}             & \multicolumn{1}{r|}{4,440 (0.8\%)}         & \href{http://knowyourmeme.com/memes/demotivational-posters}{Demotivational Posters}                            &                                      7,781 (0.8\%)  \\
\href{http://knowyourmeme.com/memes/fake-ccg-cards}{*Fake CCG Cards}         & \multicolumn{1}{r|}{4,438 (0.8\%)}       &  \href{http://knowyourmeme.com/memes/cash-me-ousside-howbow-dah}{Cash Me Ousside/Howbow Dah}                                     &      5,972 (0.6\%)                              \\ \bottomrule
\end{tabular}%
\caption{\revision{Top 10 viral memes posted on Reddit and Twitter and their respective post counts between July 2016 and July 2017 (obtained from~\cite{zannettou2018origins}). Note that entries are links to corresponding Know Your Meme entries. $^*$The meme CANNOT be detected as viral in our classifiers, see detail explanation in Section \ref{sec:twittercase}.}}
\label{tbl:top_memes}
\end{table*}

\subsection{Predicting Viral Memes on Twitter and Reddit}
\label{sec:twitterandreddit}

Our results so far are based on a dataset of viral and non-viral image memes shared on 4chan's Politically Incorrect board (/pol/).
\newrevision{Next, we want to understand if the features identified by our model are helpful in characterizing viral image memes on mainstream platforms.}
To do so, we collect an additional dataset of the top 10 memes shared on Twitter and Reddit according to previous work~\cite{zannettou2018origins}, for a total of 20 images (see Table~\ref{tbl:top_memes}).
\newrevision{We repeat our annotation process on these images, and run our models trained on 4chan data on the resulting dataset, to investigate whether the indicators of virality identified while training our model on /pol/ data are also present on the most popular image memes on Twitter and Reddit.}

\revision{Our models are able to correctly classify 19 out of the 20 most popular image memes on Twitter and Reddit as viral, indicating that our indicators of virality indeed generalize to other platforms.
In Section~\ref{sec:case-study}, we provide a more detailed discussion of some of these popular memes and of the predictors that allow our models to identify them as viral.
We also further discuss the only meme that is not correctly predicted by our model as viral, reasoning about the reasons for this misclassification.
}

\revision{Note that not all classifiers perform well when trained on 4chan data and tested on Reddit and Twitter one.
In fact, our best performing classifier for this experiment is KNN.
Tree-based classifiers, in particular, provide worse results than on the 10-fold cross validation.
This is a known limitation of these classifiers when dealing with limited training data.
In a nutshell, random forests can suffer from high variance across individual trees, and even aggregating across these individual trees, the strength of random forests in general, can lead to inconsistency~\cite{tang2018WhenRandomForestsa}}.

\begin{figure}[t]
    \centering
    \includegraphics[width=0.5\linewidth]{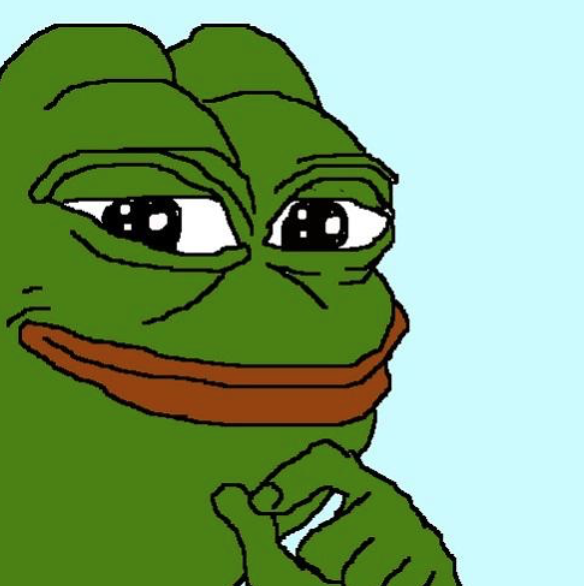}
    \caption{Smug Frog.}
    \label{fig14.png}
\end{figure}

\section{Case Studies}
\label{sec:case-study}
In this section, we discuss a number of case studies to illustrate how our models can help evaluating viral and non-viral image memes.
We first look at a viral meme (i.e., the Smug Frog), highlighting the important features that our model picks up and that can explain its virality. 
We then focus on a number of non-viral memes, looking at what features in our codebook might have had an effect.

\newrevision{Finally, we analyze the two most viral memes on Twitter and Reddit from Section~\ref{sec:twitterandreddit}, showing that the same indicators of virality that we identified for image memes on /pol/ apply to these images as well.
We also analyze the only image meme among the top Twitter and Reddit ones that our model fails to predict as viral, reasoning about why that is the case.}

\subsection{A Viral Meme: Smug Frog}
``Pepe the Frog,'' originally a comic book character created by Matt Furie, has spawned numerous derivatives and has become one of the most popular meme characters online.
In particular, polarized communities have appropriated this character, to the point that it was declared as a hate symbol by the Anti-defamation League~\cite{anderson2018counter}.
The Smug Frog (see Figure~\ref{fig14.png}) is one of the many incarnations of Pepe, which originated on 4chan in 2011, and has since been viral, with many variations posted across social media. %
Previous work which quantitatively measured the occurrence of memes~\cite{zannettou2018origins} found that this meme was prominently discussed online, appearing in 63,447 threads on 4chan, 2,197 on Twitter, 392 on Gab, and 5,968 on Reddit.

The Smug Frog presents several characteristics that our model identified as being typical of viral memes.
The illustration uses a close up scale, tightly framing the character's face.
The character presents a positive emotion and a posture, which are all important features of virality according to our model.
This shows that our model is able to correctly characterize this image meme, identifying the elements that might have contributed to its virality.

\begin{figure*}[t]
\centering
\begin{subfigure}[b]{.235\textwidth}
\includegraphics[width=1\linewidth]{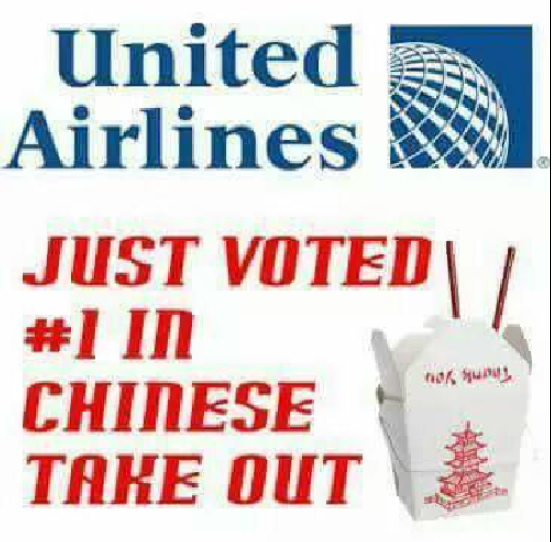}
\caption{}\label{fig/f16a}
\end{subfigure}%
~
\begin{subfigure}[b]{.235\textwidth}
\includegraphics[width=1\linewidth]{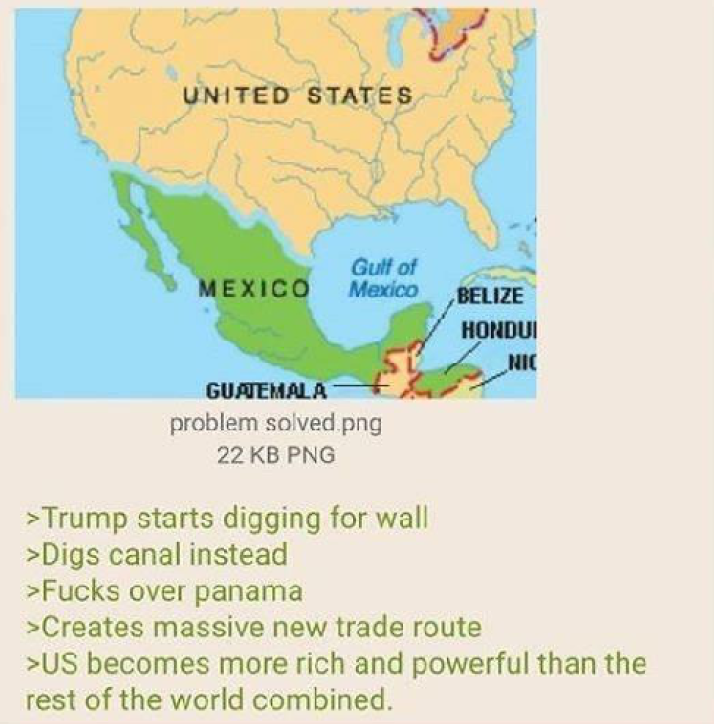}
\caption{}\label{fig/f16b}
\end{subfigure}
~
\begin{subfigure}[b]{.235\textwidth}
\includegraphics[width=1\linewidth]{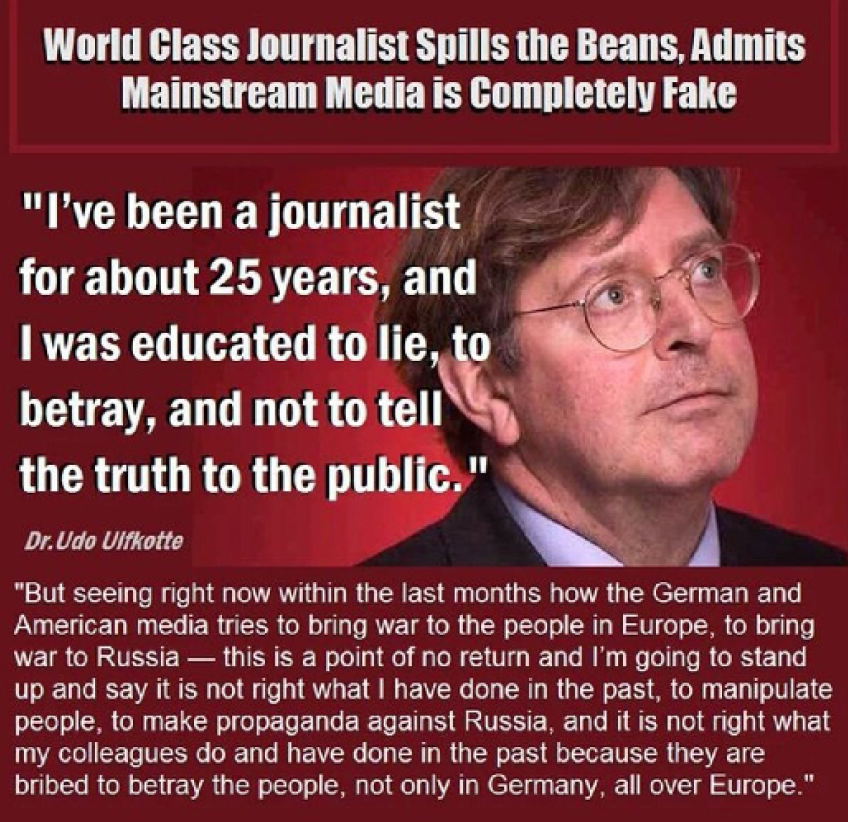}
\caption{}\label{fig/f16c}
\end{subfigure}
~
~
\begin{subfigure}[b]{.235\textwidth}
\includegraphics[width=1\linewidth]{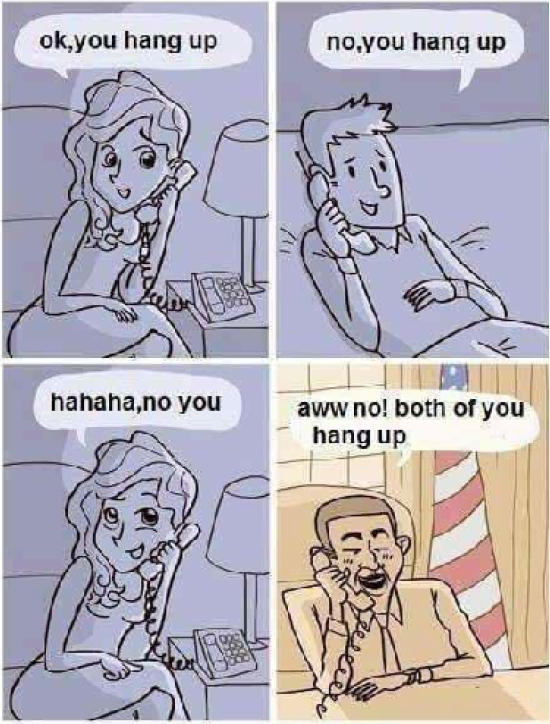}
\caption{}\label{fig/f16d}
\end{subfigure}
\caption{Non-viral memes: (a) United Airlines Passenger Removal, (b) A journalist condemn fake media, and (c) Problem solved, (d) Hang up the phone.}
\label{fig/f16}
\end{figure*}

\begin{figure*}[t]
\begin{minipage}[b]{.4\textwidth}
\centering
\begin{subfigure}[b]{.42\textwidth}
\includegraphics[width=1\linewidth]{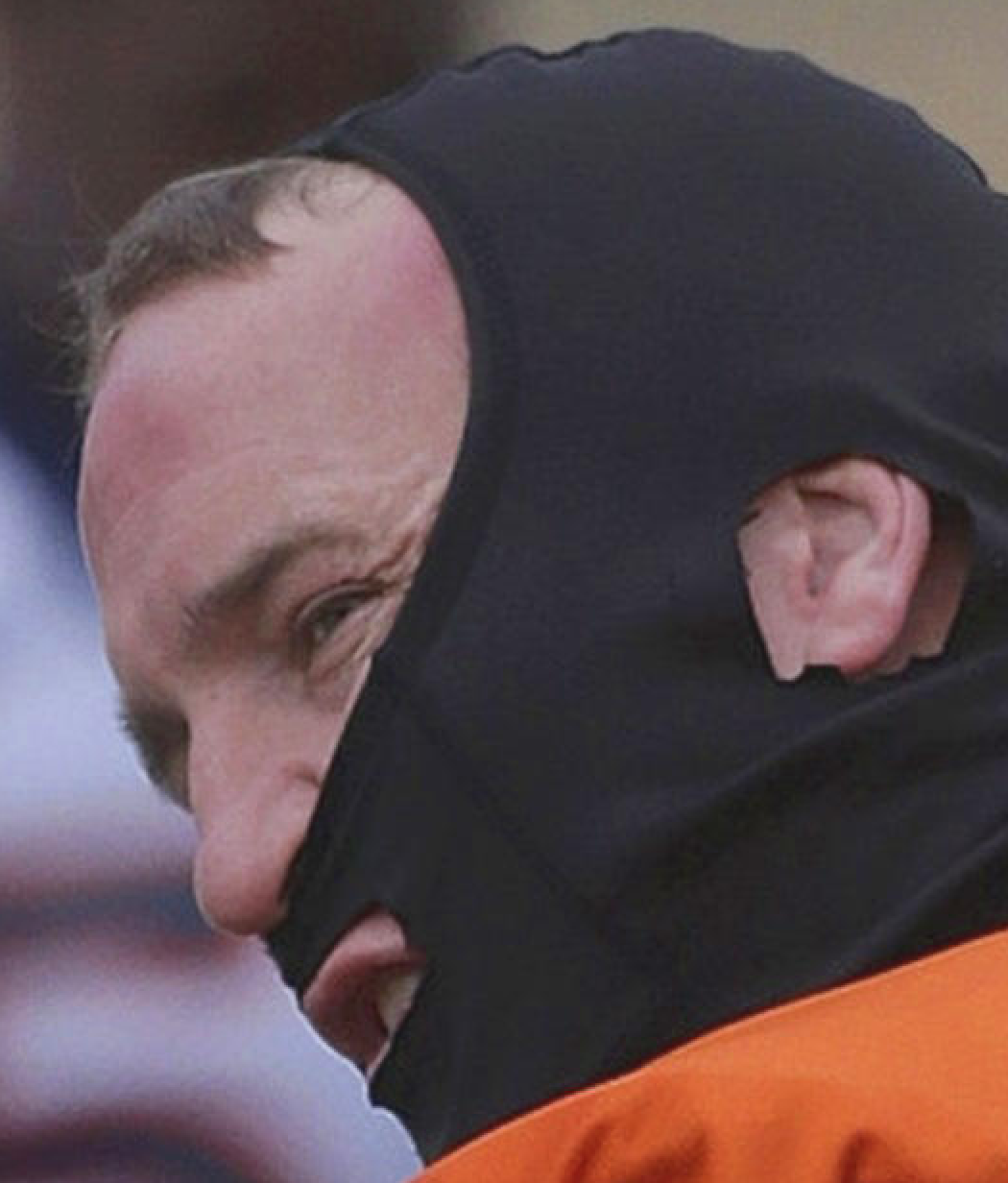}
\caption{Manning Face (1st)}\label{fig/f16a}
\end{subfigure}
~
\begin{subfigure}[b]{.54\textwidth}
\includegraphics[width=1\linewidth]{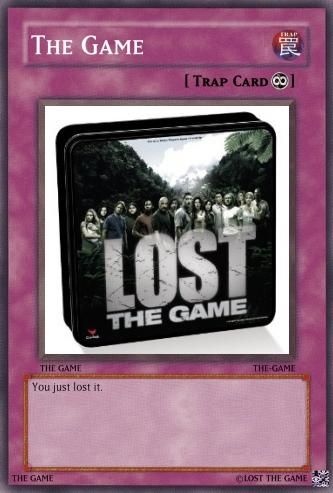}
\caption{Fake CCG Cards (10th)}\label{fig/f16b}
\end{subfigure}
\caption{\revision{Top memes on Reddit.}}
\label{fig/f50}
\end{minipage}%
\begin{minipage}[b]{.65\textwidth}
\centering
\begin{subfigure}[b]{.45\textwidth}
\includegraphics[width=1\linewidth]{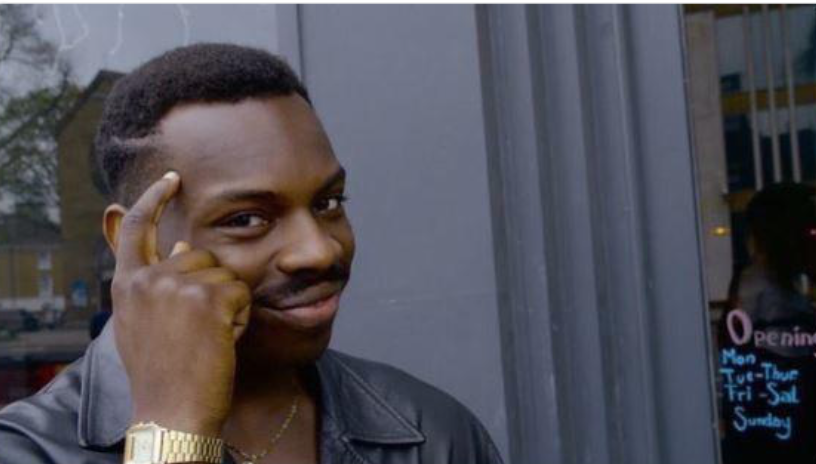}
\caption{Original Version}\label{fig/f17a}
\end{subfigure}
~
\begin{subfigure}[b]{.45\textwidth}
\includegraphics[width=1\linewidth]{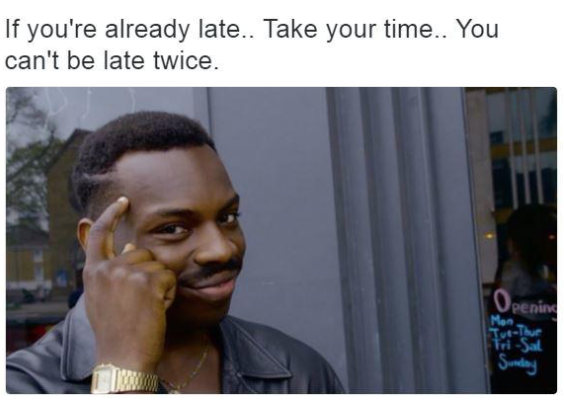}
\caption{Popular Derivative}\label{fig/f17b}
\end{subfigure}
\caption{Top meme on Twitter: Roll Safe.}
\label{fig/f17}
\end{minipage}
\end{figure*}

\subsection{Non-viral memes}

Our models have identified multiple traits that help an image meme go viral.
We also found that if an image meme lacks these characteristic traits of virality, it is unlikely to go viral. 
Consider for instance the four examples of memes that did not go viral on /pol/ displayed in Figure~\ref{fig/f16}.

At a first glance, the images lack the important indicators of virality mentioned before. 
Figures~\ref{fig/f16a} and~\ref{fig/f16b} lack characters as subjects, and therefore cannot rely on their posture or displayed emotions to bring their message across. 
While Figure~\ref{fig/f16c} and~\ref{fig/f16d} do display characters, it is not immediately clear to the viewer where to focus their attention, and both memes require the viewer to read the entire text to understand the point of the meme.
This goes against what our model has learned, that viral memes usually rely on direct visual cues and few, if any, words.
In fact, all four non-viral memes in Figure~\ref{fig/f16} rely on the user reading and understanding the rather elaborate text in order to understand the meme.

\subsection{Viral Memes on Twitter and Reddit} 
\label{sec:twittercase}
\revision{As reported in Table~\ref{tbl:top_memes}, the most shared viral meme on Reddit is Manning Face (see Figure~\ref{fig/f16a}).
According to~\cite{zannettou2018origins}, 12,540 posts on Reddit included it between 2016 and 2017, i.e., 2.2\% of all posts on Reddit.
This image meme depicts NFL quarterback Peyton Manning wearing a black hoodie, and is often used as a reply in Reddit threads as a bait-and-switch joke. %
Analyzing the meme according to our model, Manning Face presents several features that are indicative of viral memes.
It is a close up of a character presenting a dramatic facial expression.}

\revision{
The most popular meme on Twitter is Roll Safe (see Figure~\ref{fig/f17a}).
According to~\cite{zannettou2018origins}, 55,010 tweets included Roll Safe between 2016 and 2017, which represent 5.9\% of all image memes posted on Twitter.
Roll Safe was also among the top 10 popular meme on Reddit (4,523 posts, 0.8\%).}
\revision{This meme depicts actor Kayode Ewumi pointing to his temple, and is usually accompanied by witty text denoting smart thinking.
In Figure~\ref{fig/f17b}, a user captioned the meme with ``If you're already late.. take your time.. you can't be late twice.''} %
\revision{Similar to Manning Face, Roll Safe presents many of the features that we learned being indicative of viral memes: it depicts a character in a close up scale, presenting a positive facial expression and a particular posture.}

\revision{Our classifiers fails to detect one viral meme from Reddit: ``Fake CCG Card'' (see Figure~\ref{fig/f16b}), which ranks 10th on Reddit~\cite{zannettou2018origins}.
This image meme has small text, does not really use composition to attract attention, does not depict emotion, etc., and thus is a clear false negative for our model.
However, a bit more explanation about this particular meme can help reason about \emph{why} our model failed.
The ``Fake CCG Card'' meme depicts a modified version of a playing card from Collectible Card Game (CCG), e.g., Magic The Gathering or Yugioh.
This particular instance is actually a composition of the Fake CCG Card meme and another meme, ``THE GAME,'' which you lose by thinking about the phrase ``the game.''
As can be seen in Figure~\ref{fig/f16b}, by viewing this ``The Game trap card,'' you have indeed lost ``THE GAME.''
``THE GAME'' meme itself is quite a bit older than most memes, having originated in the real world in the 1990s, and while this instance is definitely an image meme, the ``meta'' aspect of ``THE GAME'' is a likely explanation of why it is not captured by our model.
}

\newrevision{Overall, these three case studies show that, although our codebook and models were developed with data from /pol/, they are helpful in characterizing memes from Twitter and Reddit too.}

\section{Discussion \& Conclusion}
\label{sec:discussion}

In this work, we studied what features are indicative of image memes going viral, across three dimensions: their composition, subjects, and intended audience.
We found that certain features like the scale of an image, the presence of characters, of facial emotions, and of poses are particularly indicative of virality.
\revision{Overall, we are confident that our work will encourage further analysis of the role of images in online discussion, online abuse, and content moderation by the CSCW community.}

In the rest of this section,  we discuss the design implications of our results as well as their limitations, also prompting future work directions.

\subsection{Design Implications} 

Our work provides evidence that aesthetic properties, grounded in art theory and neuroscience findings, have predictive power in understanding whether \emph{memes} will go viral.
This finding has clear implications with respect to marketing and outreach.
For instance, our models could be used to design more effective messaging in online ad or activism campaigns and product promotion to increase reach.
Alas, a potential worrying implication is that they could help optimize the production of harmful memes, e.g., to spread disinformation~\cite{williamsdon,zannettou2019characterizing} and hateful content~\cite{kiela2020HatefulMemesChallenge,zannettou2018origins}.
Considering that there are tools to automatically generate images with text memes~\cite{dubey2018MemeSequencerSparseMatchinga} and overall visual content~\cite{Karras_2019_CVPR}, or even so-called cheap fakes~\cite{paris2019deepfakes}, adding to the understanding of memes has potential negative consequences.
However, in this context, our models could be used by online services to prioritize moderation, for example by focusing on those memes that are more likely to go viral and therefore cause the most damage on the platform.
As recent reversals of automated-moderation systems have indicated~\cite{vincent2020YouTubeBringsBack} fully-automated, machine learning based techniques are still meaningfully inadequate compared to human moderators.
Nonetheless, human moderators need tools to even hope to combat things at Web scale; thus, we believe that integrating our findings into the toolbox that human moderators use can help prioritize and reason about the decisions they make.

\revision{Finally, we also acknowledge that designing automated tools that help determine whether pictures of individuals might become viral memes can have personal consequences, including related to one's privacy and safety.
Therefore, further interdisciplinary work is needed to inform precautionary measures as well as meaningful informative feedback for social platform operators and users alike.
}

\subsection{Limitations and Future Work}
\label{sec:limitations}

Unsurprisingly, our study is not free of limitations.
\newrevision{First of all, virality is a spectrum.
While it is easy to label images that are shared tens of thousands of times as viral and those that are only shared once or twice as non-viral, most image memes fall somewhat in between.
In this paper, we choose to focus on the most and least viral image memes shared on /pol/.
We argue that since this is the first study of indicators of virality of image memes, this choice allows us to get a reasonable understanding of whether visual features allow us to distinguish between the two classes and answer our research questions.
It is, however, possible that some of these indicators also apply to image memes that are somewhat viral, but not the most popular.
We hope that our work will encourage the CSCW community to pursue more research in this space, attempting to answer more nuanced research questions.}

While our work shows that the indicators of virality learned by our model from /pol/ image memes generalize to the top memes posted on mainstream communities like Twitter and Reddit, each online community has its own culture, and it is therefore possible that additional, community-specific indicators could more accurately predict image virality on those platforms.

As part of future work, we plan to extend our analysis to image memes posted on multiple platforms.
In particular, it would be interesting to understand how the characteristics and backgrounds of different communities influence the virality of image memes posted on them, as well as  to investigate how viewers with different cultural backgrounds (e.g., from different countries) understand the same image memes, and whether this understanding influences their decision to re-share it.

An additional limitation is that posts on /pol/ are ephemeral and get periodically deleted~\cite{hine2017kek}. 
Our definition of virality considers the number of posts shared on /pol/ that include a meme, but it is not clear whether ephemerality affects how people share memes.
We plan to investigate whether memes on platforms that are not ephemeral follow sharing patterns that are significantly different from the ones on 4chan, as well as study the use of other indicators of virality, such as number of likes and retweets.

Finally, online culture changes over time, thus, an open question is how stable our indicators are over time, and whether our model would need to be re-trained frequently.
We argue that our features are not content-dependent, but rather focus on composition, types of subjects, and audience.
Therefore, we do not expect the model to significantly change over time, in a similar way in which the composition rules for ``viral'' marketing are not changing. 
Nonetheless, we hope that future work will study how the characteristics of viral memes change over time, with the goal of identifying even more accurate indicators of virality.

\descr{Acknowledgments.}
We would like to thank the anonymous reviewers for their insightful comments that helped us improve this paper.
We also want to thank Yang Di, Yue Wang, and Yuxi Hong, who helped annotating images.
This work was supported by the National Science Foundation under Grant 1942610.

\small
\bibliographystyle{abbrv}

\end{document}